\documentclass[twocolumn,showpacs,prb]{revtex4}
\usepackage{graphicx}

\begin{document}

\title{
Electronic state around vortex in a two-band superconductor 
}

\author{Masanori Ichioka} 
\email{oka@mp.okayama-u.ac.jp}
\author{Kazushige Machida}
\affiliation{Department of Physics, Okayama University,
         Okayama 700-8530, Japan}
\author{Noriyuki Nakai} 
\affiliation{Yukawa Institute for Theoretical Physics, Kyoto University,
         Kyoto 606-8502, Japan}
\author{Predrag Miranovi\'{c}} 
\affiliation{Department of Physics, University of Montenegro, 
         Podgorica 81000, Serbia and Montenegro}

\date{\today}

\begin{abstract}
Based on the quasiclassical theory, we investigate the vortex state 
in a two-band superconductor with a small gap on 
a three dimensional Fermi surface 
and a large gap on a quasi-two dimensional one, as in ${\rm MgB_2}$. 
The field dependence of zero-energy density of states is compared 
for fields parallel and perpendicular to the $ab$ plane, 
and the anisotropy of the vortex core shape is discussed 
for a parallel field.  
The Fermi surface geometry of two-bands, combining the effect of 
the normal-like electronic state on the small gap band at high fields,  
produces characteristic behavior in the anisotropy of 
$c$- and $ab$-directions. 
\end{abstract}

\pacs{74.25.Op, 74.25.Qt, 74.25.Jb, 74.70.Ad} 


\maketitle 

\section{Introduction}
\label{sec:introduction}

Since there are many materials with a multi-band structure, 
it is important to study the contribution of the multi-band structure 
on superconductivity in order to understand the superconducting state 
in these materials.\cite{Suhl}  
Superconductivity of ${\rm MgB_2}$ 
has been well studied both experimentally and theoretically 
after its discovery.~\cite{Nagamatsu} 
Now it is recognized that ${\rm MgB_2}$ is a typical example of 
a multi-band superconductor, which has two bands and 
two superconducting gaps, i.e., 
a large superconducting gap in the $\sigma$ band and 
a small gap in the $\pi$ band.\cite{Kortus,Gonnelli,Iavarone,Tsuda,Souma}  

On the other hand, the electronic states around the vortex when applying 
a magnetic field can be a probe for the anisotropy of the superconducting 
gap and the Fermi surface structure. 
The field dependence of the density of states (DOS) $N(E=0)$ depends 
on the gap anisotropy.\cite{Volovik,IchiokaQCLd1} 
From the field orientation dependence of $N(E=0)$ 
we can obtain information of the node position in anisotropic 
superconductors.~\cite{MiranovicPRB}  
These features are examined experimentally by the specific heat or thermal 
conductivity measurements.~\cite{Park,Izawa}  

When this analysis is applied to ${\rm MgB_2}$, 
we detect the feature of two-band and Fermi surface anisotropy. 
The rapid increase in the field dependence of the DOS at low fields 
as $N(E=0)\sim H^{0.23}$ 
and the large vortex core size observed by scanning 
tunneling microscopy (STM) in ${\rm MgB_2}$ 
are understood as characteristics of two-band 
superconductivity.\cite{Yang,BouquetC,EskildsenC}
These come from the contribution 
of the small gap band, whose electronic state becomes normal-like state 
by applying a weak magnetic field, as explained by theoretical 
calculations for $H \parallel c$.\cite{Nakai,Koshelev,DahmC}  
Therefore, characters of a two-band superconductor are understood for 
$H \parallel c$. 

\begin{figure} [tbh]
\includegraphics[width=7.0cm]{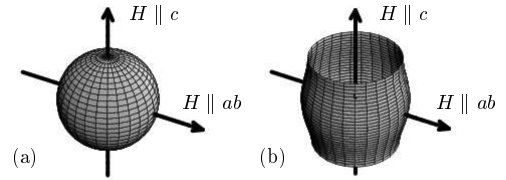} 
\caption{
Schematic view of the isotropic 3D Fermi surface (a) and 
Q2D Fermi surface (b). 
We consider the cases of two magnetic field orientations, 
$H \parallel c$ and $H \parallel ab$. 
} 
\label{fig:1}
\end{figure}

As the next step, two-band characters are examined for $H \parallel ab$, 
or by field-orientation dependence. 
In the field-orientation dependence, the contributions from the 
isotropic three dimensional (3D) Fermi surface as shown in 
Fig. \ref{fig:1}(a) are  expected to be isotropic.  
On the other hand, the quasi-two dimensional (Q2D) Fermi surface 
as shown in Fig. \ref{fig:1}(b) gives an anisotropic contribution, 
reflecting the coherence length 
ratio $\xi_{ab}/\xi_c$, which is related to the Fermi velocity  
anisotropy ratio $\langle |v_{ab}|^2 \rangle^{1/2} / 
\langle |v_{c}|^2 \rangle^{1/2} $. 
As the contributions of the $\sigma$-band with a Q2D Fermi surface and 
a $\pi$-band with a 3D Fermi surface are coupled in ${\rm MgB_2}$ and 
the dominant contribution changes depending on the field range, 
the study for $H \parallel ab$ or for 
field-orientation dependence is important 
in order to further extract the information of a two-band superconductor 
and the Fermi surface anisotropy. 

Focusing on this point, 
some experimental studies and analyses were performed. 
In the specific heat and the thermal conductivity 
measurement,\cite{Bouquet,Shibata} 
the field-dependences are compared for $H \parallel ab$ and 
$H \parallel c $, which shows that 
only a slight difference between $N(E=0,H \parallel c)$ and 
$N(E=0,H \parallel ab)$ is observed at low fields, 
while at high fields there is a pronounced anisotropy.
This result was explained due to the fact that 
the dominant isotropic contribution 
to $N(E=0)$ at low fields is coming from the $\pi$ band with 
a 3D Fermi surface, and high field anisotropic behaviors are 
due to the $\sigma$ band with a Q2D Fermi surface. 
The different temperature dependence of anisotropy ratios  
$\gamma_H=H_{c2,ab}/H_{c2,c}$ and $\gamma_\lambda=\lambda_{c}/\lambda_{ab}$ 
is also a consequence of this multi-gap (multi-band) nature 
of superconductivity in ${\rm MgB_2}$.\cite{MiranovicJPSJ}  
There were some works studying the contribution of 3D and Q2D 
Fermi surface sheets on the vortex state by quasiclassical 
theory.\cite{DahmC,Dahm,Graser} 
There, $H_{c2}$-behavior is compared for $H \parallel c$ and 
$H \parallel ab$, and the electronic states of the vortex state 
were discussed for $H \parallel c$. 
In this study, we discuss the electronic state of the 
vortex states mainly for $H \parallel ab$. 
  
For $H \parallel ab$, it is important to examine the vortex 
core shape, since the core shape is determined by the anisotropy ratio 
of the coherence length. 
For this purpose, we selfconsistently determine the profile of 
the order parameter in the vortex state. 
In this field-orientation $H \parallel ab$, 
we expect a highly anisotropic core shape 
from the Q2D Fermi surface contribution of the $\sigma$ band.  
However, the vortex core image for $H \parallel ab$ observed by STM  
shows that the vortex core is a circular shape 
at a low field.\cite{Eskildsen}    
This indicates that the anisotropic contribution of a Q2D Fermi surface 
is not clear and that the contribution of the 3D Fermi surface is dominant 
in the vortex core image of STM at this field range. 

The purpose of this paper is to investigate the electronic state 
of the vortex lattice state mainly for $H \parallel ab$ 
using quasiclassical theory in the clean limit of a two-band 
superconductor with 3D and Q2D Fermi surfaces. 
The quasiclassical theory can be applied in all range of temperatures 
and magnetic fields.  
After describing our formulation of the quasiclassical theory  
in Sec. \ref{sec:formulation}, 
we analyze the field dependence of $N(E=0)$ comparing 
two field orientations $H \parallel ab$ and $H \parallel c$ 
in Sec. \ref{sec:DOS},    
and clarify the contributions of multi-gap superconductivity and 
the Fermi surface geometries. 
In Sec. \ref{sec:LDOS},  we study 
the vortex core structure and the local density of states (LDOS), 
which are observed by STM, 
and discuss the anisotropy of the vortex core shape for  $H \parallel ab$. 
In  Sec. \ref{sec:spectrum}, 
we show the quasiparticle excitation spectrum outside vortex core 
to see the behavior of the gap edge at finite fields.  
The last section is devoted to summary and discussions. 

\section{formulation} 
\label{sec:formulation}

We consider a simplified model of a two-band system with a large 
superconducting gap band and a small gap band 
(denoted as L-band and S-band, respectively). 
For simplicity, the superconducting gap at each band is assumed 
isotropic. 
We use the following model of the 3D and Q2D Fermi 
surfaces.\cite{DahmC,Dahm,Graser} 
As shown in Fig. \ref{fig:1}(a), 
the S-band corresponding to the $\pi$-band is assumed to have 
a spherical Fermi surface, given by
$E_S({\bf k}_{\rm F})=(\hbar^2/2m)
(k_{{\rm F}x}^2+k_{{\rm F} y}^2+k_{{\rm F} z}^2)=E_{\rm F}$ 
with Fermi energy $E_{\rm F}=\hbar^2 k_{\rm F}^2/2m$. 
Fermi velocity is given by 
${\bf v}({\bf k}_{{\rm F}S})=\partial E_S/\partial {\bf k}
=v_{{\rm F}0}(k_x,k_y,k_z)/{k_{\rm F}}
=v_{{\rm F}0}(\sin\theta\cos\phi,\sin\theta\sin\phi,\cos\theta)$ 
with $0 \le \theta < \pi$ and $0 \le \phi < 2\pi$. 
As shown in Fig. \ref{fig:1}(b), 
the L-band is assumed to have a Q2D Fermi surface of 
cylinder-like shape with small ripples, given by 
$E_L({\bf k}_{\rm F})=(\hbar^2 /2m)(k_x^2+k_y^2)
-t\cos k_z=E_{\rm F}$ and Fermi velocity 
${\bf v}({\bf k}_{{\rm F}L})
=v_{{\rm F}0L} ( \cos\phi, \sin\phi, \tilde{v}_z \sin k_z) $, 
where $\tilde{v}_z$ is small 
and related to the anisotropy ratio as $1/\tilde{v}_z \sim \gamma_H$. 
In our calculation, we set $1/\tilde{v}_z = 6$ and 
$v_{{\rm F}L0} = v_{{\rm F}0}$. 

Vortex structure and electronic states are calculated by quasiclassical 
Eilenberger theory in the clean 
limit.\cite{Eilenberger,KleinJLTP,IchiokaQCLs,IchiokaQCLd1}
First, the quasiclassical Green's functions
$g({\rm i}\omega_n,{\bf k}_{{\rm F}j},{\bf r})$, 
$f({\rm i}\omega_n,{\bf k}_{{\rm F}j},{\bf r})$ and
$f^\dagger({\rm i}\omega_n,{\bf k}_{{\rm F}j},{\bf r})$ 
are calculated in a unit cell of the vortex lattice 
by solving Eilenberger equation 
\begin{eqnarray} &&
\left\{ \omega_n 
+\tilde{\bf v}({\bf k}_{{\rm F}j}) \cdot\left[
\nabla+{\rm i}{\bf A}({\bf r}) 
\right]\right\} f
=\Delta_j({\bf r})g, 
\label{eq:eil1}
\\ && 
\left\{ \omega_n 
-\tilde{\bf v}({\bf k}_{{\rm F}j}) \cdot\left[ 
\nabla-{\rm i}{\bf A}({\bf r}) 
\right]\right\} f^\dagger
=\Delta_j^\ast({\bf r})g  \quad 
\label{eq:eil2}
\end{eqnarray} 
in the so-called explosion method, 
where $g=(1-ff^\dagger)^{1/2}$, ${\rm Re} g > 0$, 
$\tilde{\bf v}({\bf k}_{{\rm F}j})
={\bf v}({\bf k}_{{\rm F}j})/v_{{\rm F}0}$, $j=L,S$,   
and Matsubara frequency $\omega_n=(2n+1)\pi T$. 
In the symmetric gauge, 
${\bf A}({\bf r})=\frac{1}{2} {\bf B} \times {\bf r} + {\bf a}({\bf r})$, 
where ${\bf B}=(0,0,B)$ is a uniform field and 
${\bf a}({\bf r})$ is related to the internal field ${\bf b}({\bf r})$ 
as ${\bf b}({\bf r})=\nabla\times {\bf a}({\bf r})$.
The unit vectors of the vortex lattice are given by 
${\bf u}_1=(a_x,0,0)$, and ${\bf u}_2=(\frac{1}{2}a_x,a_y,0)$.
The pair potential and vector potential are selfconsistently calculated by 
the relations  
\begin{eqnarray} &&
\Delta_j({\bf r})=2 T\sum_{\omega_n>0} 
\sum_{j',{\bf k}_{{\rm F}j'}}V_{jj'}
f({\rm i}\omega_n,{\bf k}_{{\rm F}j'},{\bf r}), 
\label{eq:scd}
\\ && 
{\bf J}({\bf r})
=-2T \tilde{\kappa}^{-2} 
\sum_{\omega_n>0} \sum_{j,{\bf k}_{{\rm F}j}}
{\bf v}({\bf k}_{{\rm F}j}) {\rm Im} 
g({\rm i}\omega_n,{\bf k}_{{\rm F}j},{\bf r}), \quad 
\label{eq:sca}
\end{eqnarray} 
where
${\bf J}({\bf r})=\nabla\times\nabla\times{\bf A}({\bf r})$, and  
$\tilde{\kappa}=(7 \zeta(3)/18)^{1/2}\kappa_{\rm BCS}$ 
with Rieman's zeta function $\zeta(3)$. 
We set the energy cutoff $\omega_c=20 T_{{\rm c}0}$, 
$\kappa_{\rm BCS}=30$, and 
the DOS in normal state at each Fermi surface $N_{0L}=N_{0S}=0.5N_0$. 
The integral $\sum_{j,{\bf k}_{{\rm F}j}}$ takes account of 
$N_{0j}$ and the ${\bf k}_{\rm F}$-dependent DOS 
($\propto |{\bf v}({\bf k}_{{\rm F}j})|^{-1}$) on the Fermi surface. 
The LDOS is given by 
\begin{eqnarray} &&
N_{\rm total}(E,{\bf r})=\sum_{j=L,S}N_j(E,{\bf r})
\nonumber \\ && 
=\sum_{j,{\bf k}_{{\rm F}j}} {\rm Re} g({\rm i}\omega_n 
\rightarrow E+{\rm i}\eta,{\bf k}_{{\rm F}j},{\bf r}), 
\end{eqnarray}
where $g$ is obtained by solving 
Eqs. (\ref{eq:eil1}) and (\ref{eq:eil2}) under ${\rm i}\omega_n 
\rightarrow E+{\rm i}\eta$. 
We typically use $\eta=0.01$. 
The spatial average of the LDOS gives DOS; 
\begin{eqnarray} 
N_{\rm total}(E)=\sum_{j=L,S}N_j(E)=\langle N(E,{\bf r}) \rangle_{\bf r}.
\end{eqnarray}

We assume that superconductivity in the S-band 
occurs by Cooper pair transfer $V_{LS}$ from the L-band.  
Therefore, we set the pairing interaction $V_{SS}=0$ in the S-band, 
and use $V_{SL}=0.32V_{LL}$ so that 
$\Delta_L/\Delta_S \sim 3$ at a zero field.  
Throughout this paper, temperatures, energies, lengths, magnetic fields, 
and DOS are, respectively,  measured in units of $T_{{\rm c}0}$, 
$\pi k_{\rm B}T_{{\rm c}0}$, 
$R_0=\hbar v_{{\rm F}0}/2 \pi k_{\rm B} T_{{\rm c}0}$, 
$B_0=\phi_0/2 \pi R_0^2$ and $N_0$,   
where $T_{{\rm c}0}$ is the superconducting transition temperature 
in the case $V_{LS}=V_{SS}=0$. 
Our calculation is performed at $T=0.1T_{{\rm c}0}$. 
The shape of the vortex lattice for $H \parallel c$ 
is the triangular lattice with $a_x/(2a_y/\sqrt{3})=1$. 
For $H \parallel ab$, the anisotropic ratio $a_x/(2a_y/\sqrt{3})$ 
of the vortex lattice varies from 1.2 at low fields to 
$\gamma_H \sim 6$ at high fields in 
${\rm MgB_2}$.\cite{Eskildsen,Cubitt}  
We study two cases $a_x/(2a_y/\sqrt{3})=1.5$ and 6 
for this field orientation. 

We selfconsistently calculate the order parameter 
$\Delta_L({\bf r})$, $\Delta_S({\bf r})$ and 
the vector potential ${\bf a}({\bf r})$, 
under a given vortex lattice ratio $a_x/(2a_y/\sqrt{3})$. 
We alternatively solve the Eilenberger equations 
[Eqs. (\ref{eq:eil1}) and (\ref{eq:eil2})] 
and the selfconsistent condition 
[Eqs. (\ref{eq:scd}) and (\ref{eq:sca})], 
until a sufficiently selfconsistent solution is obtained. 
As an initial state of the calculation, we use the Abrikosov 
solution of the lowest Landau level for $\Delta_L({\bf r})$, 
and set $\Delta_S({\bf r})=0$ and ${\bf a}({\bf r})=0$. 
By this selfconsistent calculation, we can properly estimate 
the size and shape of the vortex core. 
In the Abrikosov solution, vortex core size 
is always proportional to the inter-vortex distance. 
In our selfconsistent method, the vortex core size remains to be 
in the order of the coherence length even when the inter-vortex 
distance becomes large at low fields. 
This proper treatment of the vortex core size is necessary 
to correctly estimate the low energy excitations of the 
quasiparticles especially at low fields. 
The selfconsistently obtained  ${\bf a}({\bf r})$ does 
not give a significant contribution on the electronic state, 
since we consider the case of high-$\kappa$.

\section{field dependence of zero-energy density of states} 
\label{sec:DOS}

\begin{figure} [tbh]
\includegraphics[width=7.0cm]{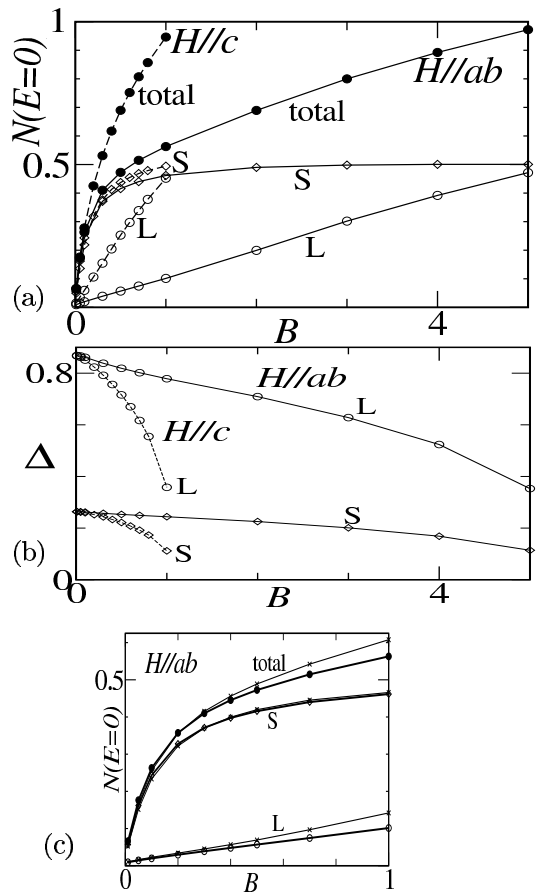} 
\caption{
(a) 
Field dependence of zero-energy DOS $N_{\rm total}=N_L+N_S$ ($\bullet$) 
for $H \parallel ab$ (solid lines) and $H \parallel c$ (dashed lines).
Large-gap band contributions $N_L$ ($\circ$) and 
small-gap band contributions $N_S$ ($\diamond$) are also presented. 
Vortex lattice ratio $a_x/(2a_y/\sqrt{3})=6$ for 
$H \parallel ab$ and 1 for $H \parallel c$. 
(b) 
Field dependence of maximum pair potential amplitudes 
$|\Delta_L|$ ($\circ$) and $|\Delta_S|$ ($\diamond$) 
in the vortex lattice state  
for $H \parallel ab$ (solid lines) and $H \parallel c$ (dashed lines).
(c) 
Field dependence of zero-energy 
DOS for $a_x/(2a_y/\sqrt{3})=1.5$ (thin lines with $\times$) 
and 6 (thick lines) for $H \parallel ab$ at low fields. 
} 
\label{fig:2}
\end{figure}

At low fields, low energy quasiparticles are localized around 
vortex cores in full-gap superconductors. 
With increasing field, 
the distribution of low energy quasiparticles around a vortex core overlaps  
with that coming from neighbor vortex cores, since inter-vortex 
distance decreases.\cite{IchiokaQCLs} 
Therefore, low energy quasiparticles extend outside of the vortex core. 
The zero-energy total DOS $N_{\rm total}(E=0)$ is the spatial 
average of these low-energy quasiparticle distributions. 
Field dependence of total DOS $N_{\rm total}(E=0)$ and 
each band contribution $N_L(E=0)$ and $N_S(E=0)$ 
are shown in Fig. \ref{fig:2}(a)  
both for $H \parallel ab$ and $H \parallel c$. 
The DOS on the S-band increases rapidly at low fields, 
and saturates to normal state value $0.5 N_0$ at higher fields. 
The increase of total DOS at high fields is due to the L-band contribution. 
These are consistent with previous 
calculations.~\cite{Nakai,Koshelev,DahmC} 

In this study, 
we compare the curves for  $H \parallel ab$ and $H \parallel c$. 
The rapidly increasing $S$ band contribution $N_S$ is almost the same 
for both orientations at low fields, 
reflecting the 3D isotropic Fermi surface shape. 
Slowly increasing the L-band contribution $N_L$ 
shows a large difference, reflecting the anisotropy of $H_{c2}$ 
dominantly coming from the Q2D Fermi surface. 
As a result, $N_{\rm total}$ is almost isotropic at low fields, 
and largely anisotropic at higher fields, 
which is consistent with experimental observations.~\cite{Bouquet,Shibata} 

Figure \ref{fig:2}(b) indicates that, while electronic states in the S-band 
are normal-like state at high fields, the pair potential $\Delta_S$ 
in the S-band persists up to $H_{{\rm c}2}$  
with a gap ratio $|\Delta_L|/|\Delta_S|$ relatively unchanged, 
as well as that $\Delta_S$ persists up to $T_{\rm c}$ 
at $H=0$.\cite{Gonnelli,Iavarone,Tsuda} 
This is easy to understand since superconductivity in the S-band is 
induced by the Cooper pair transfer from the L-band. 
As long as we have the pair potential $|\Delta_L|$ in the L-band, 
the induced pair potential in $S$-band is non-vanishing. 
This behavior is similar both for $H \parallel c$ and $H \parallel ab$. 

It is necessary to see how DOS depends on the anisotropic ratio 
$a_x/(2a_y/\sqrt{3})$ for a unit cell of the vortex lattice, 
which is given in our calculation. 
We plot the DOS for $a_x/(2a_y/\sqrt{3})$=6 and 1.5 in Fig. \ref{fig:2}(c). 
At low fields, DOS does not significantly depend 
on the shape of the unit cell, 
because low-energy electrons are localized around vortex cores. 
Therefore, our numerical results do not significantly depend  
on the delicate tuning of the vortex lattice ratio 
$a_x/(2a_y/\sqrt{3})$ at low fields. 
At higher fields, where the low-energy quasiparticles around vortices 
overlap each other, there appears a deviation depending 
on the vortex lattice shape. 
At these fields, however, 
the equilibrium vortex lattice in ${\rm MgB_2}$ 
is a distorted hexagonal with 
$a_x/(2a_y/\sqrt{3})\sim \gamma_H \sim 6$.\cite{Cubitt}  
Therefore, it is reasonable to study the vortex structure 
with $a_x/(2a_y/\sqrt{3})=6$ at high fields. 

\begin{figure} [tbh]
\includegraphics[width=6.0cm]{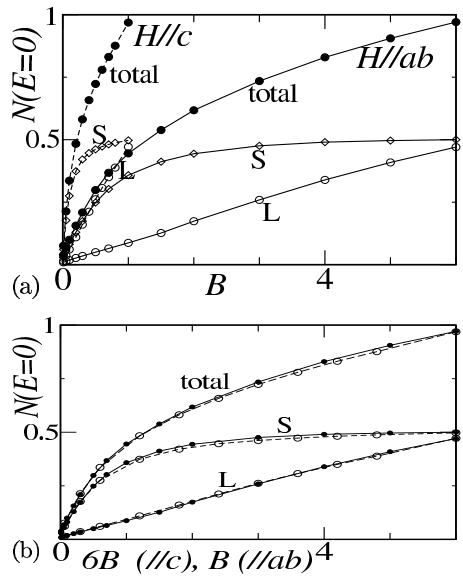} 
\caption{
(a) 
Field dependence of DOS $N_{\rm total}$ ($\bullet$), 
$N_L$ ($\circ$) and $N_S$ ($\diamond$) 
for $H \parallel ab$ (solid lines) and $H \parallel c$ (dashed lines) 
when both bands have Q2D Fermi surfaces with $1/\tilde{v}_z=6$. 
$a_x/(2a_y/\sqrt{3})=6$.  
(b) 
Field dependence of (a) is replotted as a function of $6B$ 
for $H \parallel c$  ($B$ for $H \parallel ab$), 
as $\gamma_H \sim 6$. 
} 
\label{fig:3}
\end{figure}

To show an example that the field-dependence of $N_{\rm total}$ 
is changed by the Fermi surface geometry
in a two-band superconductor, 
we also calculate the case 
when both L- and S-bands have Q2D cylindrical Fermi surfaces  
with $1/\tilde{v}_z=6$, for comparison.  
In this case, as shown in Fig. \ref{fig:3}(a), 
field-dependences of $N_{\rm total}$ are anisotropic between 
$H \parallel ab$ and $H \parallel c$ both at low and high field,  
because S-band contributions at low fields are also anisotropic 
due to the Q2D Fermi surface contribution. 
When all bands have the same Fermi surface geometry as in this case, 
we can expect that $H$-dependences of $N_{\rm total}(E=0)$ are 
roughly scaled by $H_{\rm c2}$ anisotropy. 
In Fig. \ref{fig:3}(b), 
the field dependences of zero-energy DOS 
are replotted as a function of $6B$ for $H \parallel c$  
($B$ for $H \parallel ab$) as $\gamma_H \sim 6$. 
There, calculated data of $N_{\rm total}$ (and also $N_L$, $N_S$) 
for $H \parallel ab$ and $H \parallel c$ are on the same curve, 
i.e., field dependences are well scaled by $H_{\rm c2}$ anisotropy. 
On the other hand, the field dependences in Fig. \ref{fig:2}(a) 
are not scaled by the $H_{c2}$ ratio, because the two bands have 
different geometry, i.e., 3D and Q2D Fermi surfaces.  

\begin{figure} [tbh]
\includegraphics[width=7.5cm]{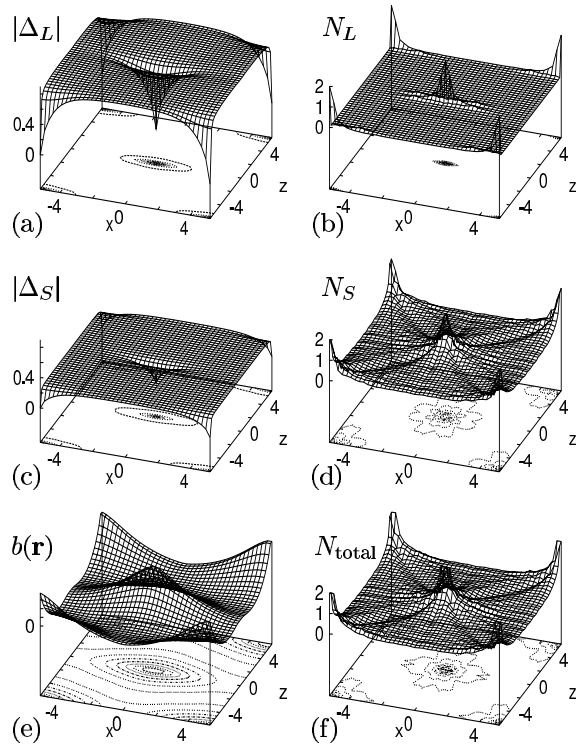} 
\caption{
Spatial structure of vortex lattice state for  $H \parallel ab$. 
$B=0.1$ and $a_x/(2a_y/\sqrt{3})=1.5$.  
The vortex centers are located in the middle and at the four corners 
of the figure. 
The pair potential amplitude $|\Delta_L({\bf r})|$ (a) and  
$|\Delta_S({\bf r})|$ (c). 
Internal field $b({\bf r})$ (e).  
Zero-energy LDOS $N_{\rm total}(E=0,{\bf r})$ (f) 
and each band contribution $N_L(E=0,{\bf r})$ (b) 
and $N_S(E=0,{\bf r})$ (d). 
The peaks are truncated at $N=2N_0$.
} 
\label{fig:4}
\end{figure} 

\section{vortex core structure for parallel fields} 
\label{sec:LDOS}

We study the vortex core structure and the LDOS around the vortex 
for $H \parallel ab$ in order to see how the anisotropy of the vortex 
core shape is affected by 3D or Q2D Fermi surfaces. 
Figure \ref{fig:4} shows the vortex structure 
for $H \parallel ab$ at $B=0.1$. 
In this low field case, 
we use the vortex lattice ratio $a_x/(2a_y/\sqrt{3})=1.5$ 
for the unit cell of the vortex lattice. 
Looking at the amplitude of the selfconsistently 
calculated pair potential $|\Delta_L({\bf r})|$ [(a)] 
and $|\Delta_S({\bf r})|$ [(c)] under the given vortex lattice ratio, 
one can notice that the vortex core shape is highly anisotropic 
due to the effect of the Q2D Fermi surface. 
The anisotropy of internal field $b({\bf r})$ [(e)] is not so large. 
The LDOS on the L-band, $N_L(E=0,{\bf r})$ [(b)], 
is highly anisotropic, reflecting anisotropy of $|\Delta_L({\bf r})|$.  
However, the LDOS on S-band, $N_S(E=0,{\bf r})$ [(d)],  
shows isotropic distribution reflecting 3D Fermi surface. 
The LDOS around the vortex core is broad on the S-band, 
as in the case $H \parallel c$.\cite{EskildsenC,Nakai,Koshelev}  
Since the S-band contribution is dominant at this low field, 
as discussed in Fig. \ref{fig:2}(a), 
the total LDOS $N_{\rm total}(E=0,{\bf r})$ [(f)] 
shows the isotropic vortex core shape, reflecting $N_S(E=0,{\bf r})$. 
This corresponds to an almost isotropic vortex core shape 
observed by STM.\cite{Eskildsen}  
The highly anisotropic L-band contribution is masked 
by the S-band contribution. 
Therefore, while the pair potential around the vortex core is highly 
anisotropic reflecting the Q2D Fermi surface, the dominant contribution 
of the S-band with the 3D Fermi surface gives an isotropic vortex core 
image in the quasiparticle excitations observed by STM.

\begin{figure} [tbh]
\includegraphics[width=7.5cm]{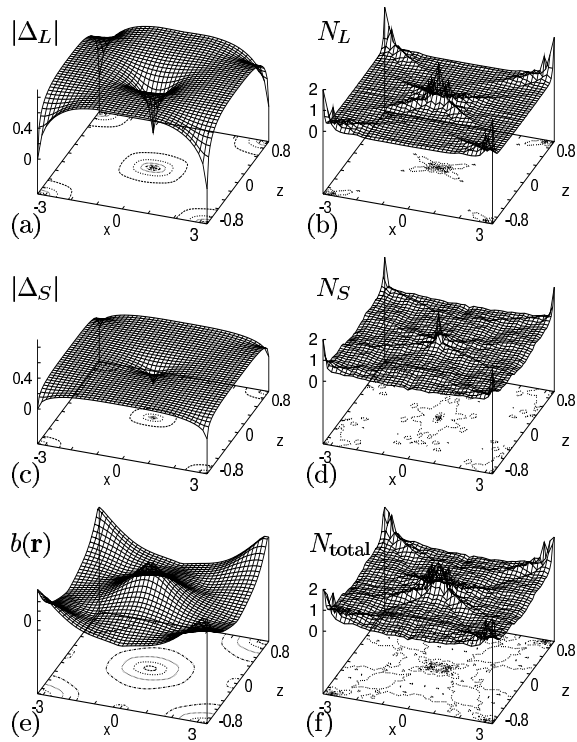} 
\caption{
The same as Fig. \ref{fig:4}, but $B=1$ and $a_x/(2a_y/\sqrt{3})=6$.  
It is noted that scales of $x$ and $z$ are anisotropic in this 
figure of a highly distorted hexagonal vortex lattice case 
at high fields. 
} 
\label{fig:5}
\end{figure} 

Figure \ref{fig:5} shows the vortex structure at a higher field  
$B=1$, where we use the vortex lattice ratio of a highly distorted hexagonal, 
$a_x/(2a_y/\sqrt{3})=6$. 
The difference of overall views in Figs. \ref{fig:4} and \ref{fig:5} 
also comes from the deformation of the vortex lattice. 
While the vortex core structures of 
$|\Delta_L({\bf r})|$ [(a)],   $|\Delta_S({\bf r})|$ [(c)] 
and $N_L(E=0,{\bf r})$ [(b)] are similar to those in  Fig. \ref{fig:4}, 
they are seen as if they are isotropic in this figure scaled 
by the coherence length ratio $\gamma_H \sim 6$. 
That is, when we see the pair potential amplitude and $N_L(E=0,{\bf r})$, 
the vortex core structure is highly anisotropic, 
reflecting the Q2D Fermi surface,  both at low and high fields. 
The anisotropy around the vortex in $b({\bf r})$ increases 
with raising the field, 
and saturates to that of $\gamma_H$ as is seen in Fig. \ref{fig:5}(e). 
By the effect of increasing the field, 
$N_S(E=0,{\bf r})$ [(d)] is almost flat 
and slightly enhanced at the vortex center. 
Therefore, the vortex core shape in the total LDOS 
$N_{\rm total}(E=0,{\bf r})$ [(f)] reflects the spatial structure 
of $N_L(E=0,{\bf r})$ at higher fields.

\section{local spectrum far from vortex core} 
\label{sec:spectrum}

As mentioned above, while both pair potentials $|\Delta_L|$ and 
$|\Delta_S|$ survive up to $H_{\rm c2}$, 
the electronic state in the S-band becomes a normal-like state 
at high fields. 
The field range of this normal-like state is wider for $H \parallel ab$. 
Therefore, it is interesting to see how the superconducting gap edges at 
$|\Delta_L|$ and $|\Delta_S|$ in the quasiparticle excitation spectrum 
are smeared by the quasiparticle excitations in the vortex states. 

\begin{figure} [tbh]
\vspace{0.5cm}
\includegraphics[width=6.5cm]{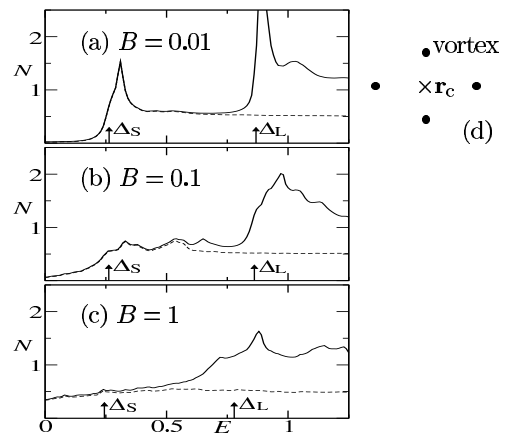} 
\caption{
Local spectrum $N_{\rm total}(E,{\bf r}_{\rm c})$ [solid lines] 
and the S-band contribution  $N_S(E,{\bf r}_{\rm c})$ [dashed lines] 
for $H \parallel ab$ at the midpoint ${\bf r}_{\rm c}$ 
of the surrounding four vortices as shown in (d). 
$B=0.01$ (a), 0.1 (b) and 1 (c). 
$a_x/(2a_y/\sqrt{3})=1.5$ ($B=0.01$, 0.1) and 6 ($B=1$).  
Arrows in the figure indicate $|\Delta_L|$ and $|\Delta_S|$ 
at each field, presented in Fig. \ref{fig:2}(b).
} 
\label{fig:6}
\end{figure} 

The local spectrum $N_{\rm total}(E,{\bf r}_{\rm c})$ 
is shown in Fig. \ref{fig:6} 
at the point ${\bf r}_{\rm c}$ far from vortex cores, 
where the superconducting gap structure in the spectrum 
may survive up to higher fields, compared with other points. 
At a low field $B=0.01$ [Fig. \ref{fig:6}(a)], 
S-band and L-band contributions, respectively, have clear gap edges 
at $E\sim \Delta_S$ and $\Delta_L$. 
With increasing $B$ [Fig. \ref{fig:6}(b)], 
low energy DOS extending from the vortex core grows up within the gap edge, 
and the gap edge is smeared. 
The smearing is eminent for the lower excitation gap of $\Delta_S$, 
whose excitation gap structure comes from the S-band 
contribution $N_S(E,{\bf r}_{\rm c})$. 
At high field $B=1$ [Fig. \ref{fig:6}(c)],  
$N_S(E,{\bf r}_{\rm c})$ is almost flat, i.e., normal-like state. 
While there are some attempts to observe the field dependence of 
$\Delta_S$ by STM,\cite{Bugoslavsky,GonnelliH} 
the excitation gap of $\Delta_S$ becomes difficult 
to be observed at the field range of a normal-like state S-band state. 

\section{summary and discussions} 
\label{sec:summary}

Based on the quasiclassical theory, 
the electronic structure in the vortex state was studied 
in a two-band superconductor with the 3D and Q2D Fermi surfaces, 
mainly for $H \parallel ab$. 
To demonstrate the appearance of the effect due to the 
geometry of the 3D and Q2D Fermi surfaces, 
we compared the field dependence of zero-energy DOS 
for $H \parallel ab$ and $H \parallel c$, 
and analyzed the anisotropy of the vortex core structure 
for $H \parallel ab$. 
The 3D and Q2D Fermi surfaces, respectively, 
introduce the isotropic and anisotropic contribution, 
when $c$- and $ab$- orientations are compared. 
At low (high) fields, dominant contribution to the change of 
the total DOS is coming from the small (large) gap band 
which has the isotropic 3D (anisotropic Q2D) Fermi surface. 
Therefore, the field-dependence of zero-energy DOS shows  
isotropic (anisotropic) behavior, when the field-orientation is changed. 
This reproduces the results of specific heat and thermal 
conductivity measurements.~\cite{Bouquet,Shibata} 
As for the vortex core anisotropy, 
the pair potential around the vortex core is highly 
anisotropic reflecting the Q2D Fermi surface. 
However, the dominant contribution of the low-energy qausiparticle state,  
coming from the S-band with a 3D Fermi surface, 
gives an isotropic vortex core image at low fields. 
This may be a explanation of the almost circular vortex 
core image by STM in ${\rm MgB_2}$ even when 
$H \parallel  ab$.~\cite{Eskildsen}  

As is seen in DOS, LDOS and spectrum, the electronic state 
on the S-band is normal-like state in a wide field range at higher fields. 
However, this does not mean that the pair potential 
$|\Delta_S|$ on the S-band vanishes in this field region. 
Quite contrary, $|\Delta_S|$ persists up to $H_{{\rm c}2}$ 
since superconductivity in the S-band is 
induced by he Cooper pair transfer from the L-band. 
The origin of a normal-like electronic state on the S-band is the 
contribution of low energy excitations by a supercurrent around 
vortex cores, coupling with the vortex core state of the quasiparticles. 
With an increasing magnetic field, the low energy excitations are enhanced, 
and those low energy quasiparticles around the vortex are 
delocalized giving rise to large LDOS between vortices. 
For the small gap in the S-band, the low field gives enough excitations 
to smear the gap structure of $\Delta_S$, 
resulting in a normal-like electronic state. 
We demonstrated that ${\rm MgB_2}$ shows characteristic 
behavior in the anisotropy of $c$- and $ab$-axis directions, 
by coupling of the effect of these normal-like electronic states 
in the S-band at high fields and the effect of the Fermi surface geometry 
(3D and Q2D Fermi surfaces) in a two-band superconductor. 


\end{document}